# Acknowledging user requirements for accuracy in computational chemistry benchmarks

Andreas Savin[1*] and Pascal Pernot[2**]

[1] Laboratoire de Chimie Théorique, CNRS and UPMC Université Paris 06, Sorbonne Universités, 75252 Paris, France;
[*] Electronic Address: Andreas.Savin@lct.jussieu.fr

[2] Institut de Chimie Physique, UMR8000, CNRS, Université Paris-Saclay, 91405 Orsay, France
[**] Electronic Address: Pascal.Pernot@universite-paris-saclay.fr

Dedicated to Professor Juri Grin on the Occasion of his 65th Birthday

Abstract

Computational chemistry has become an important complement to experimental measurements. In order to choose among the multitude of the existing approximations, it is common to use benchmark data sets, and to issue recommendations based on numbers such as mean absolute errors. We argue, using as an example band gaps calculated with density functional approximations, that a more careful study of the benchmark data is needed, stressing that the user's requirements play a role in the choice of an appropriate method. We also appeal to those who measure data capable of being used as a reference, to publish error estimates. We show how the latter can affect the judgment of approximations used in computational chemistry.

# Introduction

Dirac's 1929 statement[1]:

*The underlying physical laws necessary for the mathematical theory of a large part of physics and the whole of chemistry are thus completely known, and the difficulty is only that the exact application of these laws leads to equations much too complicated to be soluble. It therefore becomes desirable that approximate practical methods of applying quantum mechanics should be developed, which can lead to an explanation of the main features of complex atomic systems without too much computation.*

is still valid. To assert the validity of an approximate method, typically a density functional approximation, validation through reference data is needed. However, the required accuracy depends on the user and the application. Furthermore, the reference data are not exempt of errors. These limits are important, but often ignored, both for the reference data, and the requirements put on an approximation. Or, by convenience, some standard is set. For example, one sets a "chemical accuracy" for thermochemistry to 1 kcal/mol, but the accuracy of the reference data, or the one required in computational chemistry, may be larger or smaller.

Nowadays, a large amount of data is generated, and statistical methods are used to replace personal experience. Benchmarks are produced to recommend one or several specific approximations. We present in this paper some aspects of the impact of uncertainty on these benchmarks. We would like to stress that the aim of this paper is solely to exemplify this latter point, and by no means to recommend or to denigrate any density functional or benchmark.

# Example and notations

To exemplify our opinions, we select a benchmarking of band gaps presented by Borlido *et al.*[2]. From the fifteen approximations tested in this paper, we select only two:

- LDA, the local density approximation, notorious for underestimating band gaps[3], and

- HSE06, the more expensive approximation by Heyd, Scuseria, and Ernzerhof[4] reparameterized in 2006[5], that has become a standard for more reliable band gaps.

Figure 1 presents the errors in the calculated band gaps for the 471 systems of the benchmark set. It confirms what we already know: LDA gaps are too small, while HSE06 values are much better.

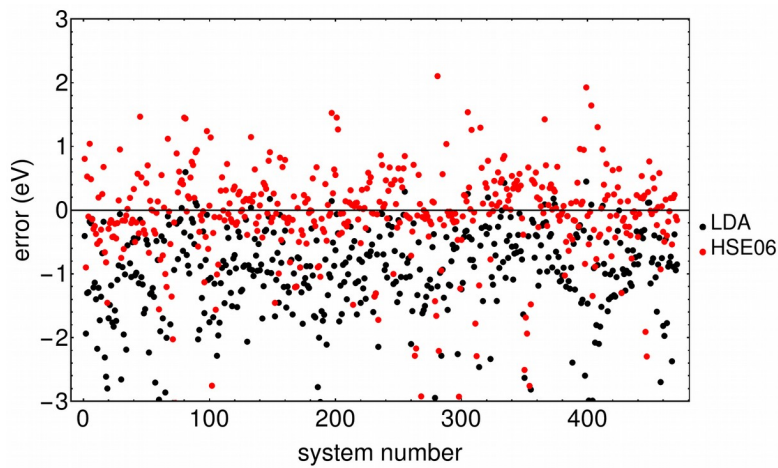

*Figure 1. Calculated minus experimental band gap (LDA black points, HSE06 red points) for the 471 systems of the benchmark of Borlido et al.*[2]

Let us introduce some notation. The number of systems studied is $N=471$. Let the error for system $i$ in method $m$ be $e_{m,i}$. The mean error for method $m$ is estimated by

$$\mu_m = 1/N \sum_{i=1,N} e_{m,i} \tag{1}$$

The standard deviation is estimated by

$$\sigma_m = \{1/(N-1) \sum_{i=1,N} [e_{m,i}-\mu_m]^2 \}^{1/2} \qquad (2)$$

The mean of the errors synthesizes what is seen in Fig. 1: $\mu_{LDA}$ = -1.2 eV, $\mu_{HSE06}$ = -0.1 eV. However, we also notice in this figure that the spread of errors is not so different. This is confirmed by $\sigma_{LDA}$ = 1.1 eV and $\sigma_{HSE06}$ = 0.8 eV.

## Systematic errors correction

In many cases, one does not need to know the exact value of a property, but how this value changes from one system to another. For example, measurements exist for a given system, and one needs to know what happens if this system is modified. Or one needs only to know how the quantity of interest varies in a class of systems (to follow a trend). In such cases, a systematic error is not important. Shifting the error sets by a constant does not change these observations.

Let us thus shift the calculated values by the mean error for the corresponding method (cf. Fig. 2). Now both methods have zero mean error, so-called centered errors. The difference between the methods is less important now, because the spread of the errors is comparable. This is reflected on the values of σ given above, both of the order of 1 eV. The maximal errors within the test set are in both cases of the order of 2 eV.

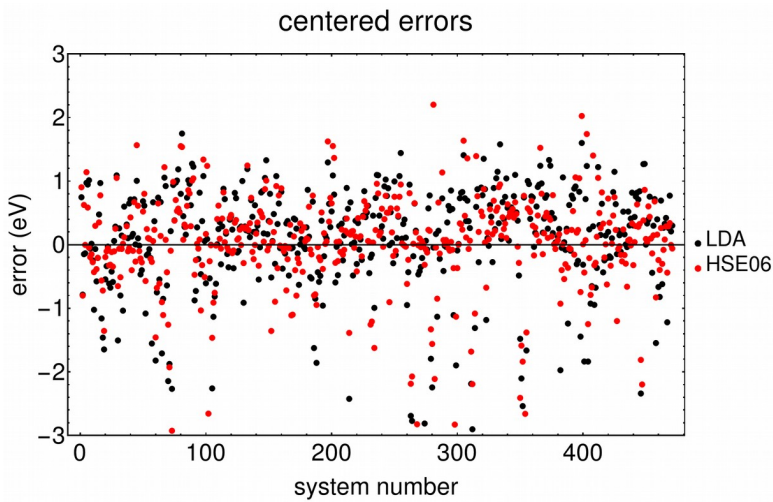

Figure 2. Calculated minus experimental band gap (LDA black points, HSE06 red points) for the 471 systems of the benchmark of Borlido et al.[2]. Centered errors.

A warning should be issued for the case of shifting the errors, as this can bring some unwanted effects. If the shift reduces the band gaps by a constant, it is possible that those smaller than the shift become negative (that is absurd). In the examples above, the band gaps were shifted upwards. However, this case also deform the results, *e.g.*, transforming a metal into a semiconductor. In order to get around these problems, one could exclude systems with small band gaps from the shift. Or, instead of shifting, one could use the benchmark data to produce scaling factors.

## Cumulative distribution functions

As the errors can become quite large, it is necessary to quantify how likely it is to obtain such large errors. For this, let us look at the cumulative distribution functions (CDFs). In probability theory, the CDF is defined as the probability that a variable $X$ takes a value smaller than $x$. For us this can be described by the fraction of systems having an error smaller than $x$. We consider from now on only absolute values of the errors $|e_{m,i}|=X$, and some acceptable upper limit of the error, that has to be defined by the user. Multiplying this number by 100 give the percentage of systems having errors smaller than $x$. Fig. 3 shows the CDFs for LDA and HSE06.

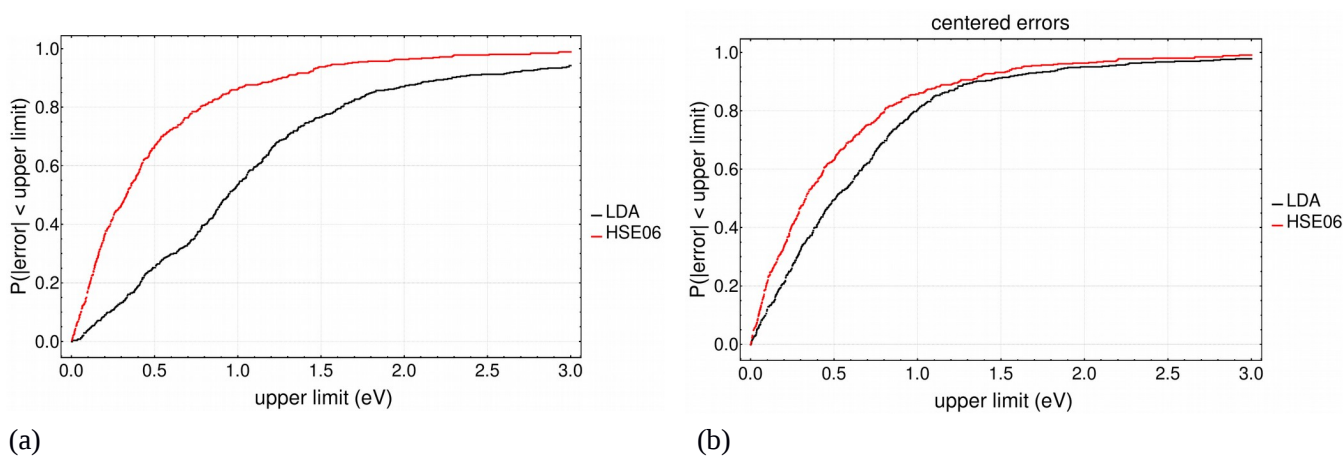

(a)  (b)

Figure 3. CDFs (fraction of systems having absolute errors having an error smaller than some preset upper limit; LDA in black, HSE06 in red) based upon the 471 systems of the benchmark of Borlido et al.[2]: (a) errors; (b) centered errors.

Let us suppose that an acceptable absolute error is of 0.5 eV. We see in Fig. 3(a) that for LDA, only a quarter of the systems of the benchmark set satisfy this condition, while for HSE06 about 2/3 of the systems satisfy it.

Let us now use the mean errors to correct the computed values (as done for studying trends). The LDA curve gets much closer to the HSE06 curve (Fig. 3(b)).

Using, as above, an upper limit for the acceptable error at 0.5 eV, we see that the percentage of systems where the error is acceptable is practically unchanged with HSE06, while it increases to about 50% with LDA after correcting for systematic errors. If we increase the tolerance to 1 eV, the "unacceptable" results are about 20% of the cases, both for LDA and HSE06.

Of course, one does not know *a priori* what an acceptable error is. The benchmark does not specify it, but the user. Or seen differently, the user has to define a risk: in which percentage of his future calculations can he accept to be wrong?

## Pairwise comparison

From the comparison of the CDFs, one might get the impression that HSE06 is always better than LDA, even after making the shift. However, this must not be the case: the curves above show only an overall behavior, not that for a specific system.

For this, let us consider the differences of the absolute errors of the two methods, system by system,

$$\Delta_i = |e_{\text{HSE06},i}| - |e_{\text{LDA},i}| \qquad (3)$$

For systems where the difference is negative, HSE06 is better than LDA. If it is positive, LDA is better than HSE06. The fraction of systems having negative difference gives the systematic improvement probability (SIP)[6] of HSE06 over LDA. Please notice that SIP = 0.5 would mean that there is equal probability of improvement or worsening by changing from LDA to HSE06.

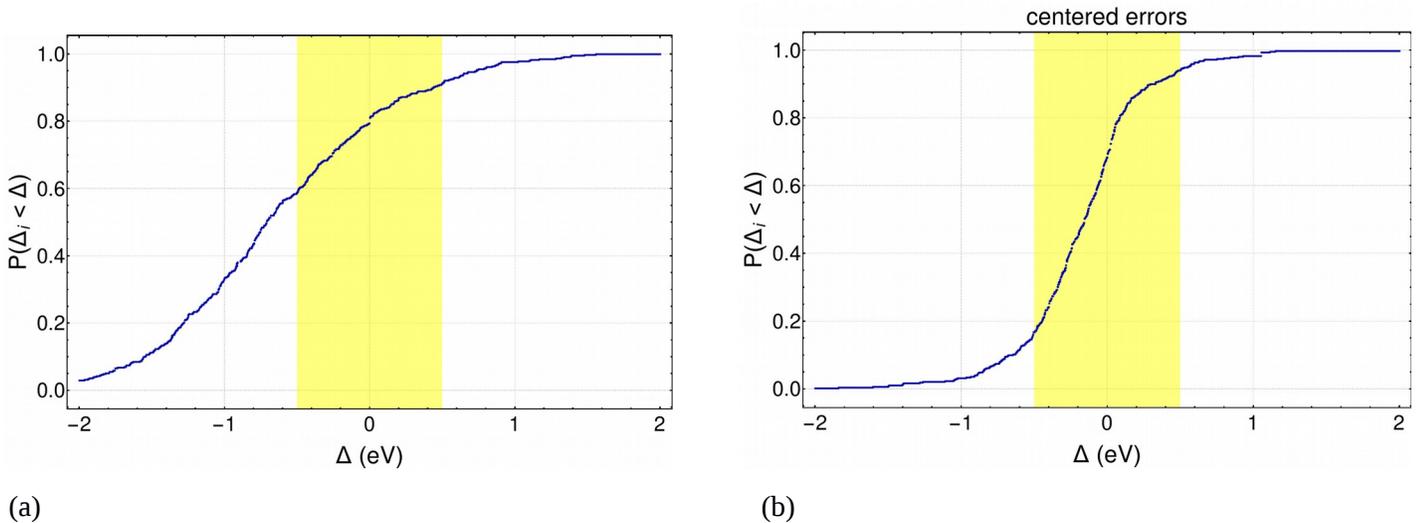

(a)          (b)

*Figure 4. CDFs (fraction of systems having $\Delta_i = |e_{\text{HSE06},i}| - |e_{\text{LDA},i}|$ smaller than some prescribed value, $\Delta$), based upon the benchmark of Borlido, et al.[2]: (a) errors; (b) centered errors. The yellow area marks the "acceptable" error margin of 0.5 eV.*

In Fig. 4(a) one can see this fraction of systems having $\Delta_i$ smaller than some prescribed value, $\Delta$. From its value at $\Delta=0$, we obtain the SIP. It is close to 0.8. Thus HSE06 is superior to LDA in 80 per cent of the cases, not always. One can look at different values of $\Delta$. If one accepts that differences below 0.5 eV are not significant, HSE06 is superior to LDA in about 60 per cent of the cases, and inferior in about 100-90=10 per cent of the cases.

We can consider what happens after eliminating the systematic errors (after correction by the mean). We see in Fig. 4(b) that HSE06 is now superior to LDA in about 2/3 of the cases.

If one considers that differences smaller than 0.5 eV are not significant, after the correction for systematic errors, only one system in 6 yields superior results when HSE06 is used, and the chance that

LDA is superior to HSE06 is three times smaller; in almost 80 per cent of the cases, the difference is considered irrelevant.

## Importance of reference data uncertainty

Up to now, it was assumed that the reference data have no uncertainty. Let us now assume that they have one, identical for all systems, characterized by a standard deviation σ.

Introducing this uncertainty, the SIP spreads out. This effect is seen in Fig. 5, where the distribution of SIP due to random reference data errors is shown. (It is estimated from generated sets of data produced by adding to the reference data randomly generated errors coming from a normal distribution centered around 0 with different values for standard deviation σ.)

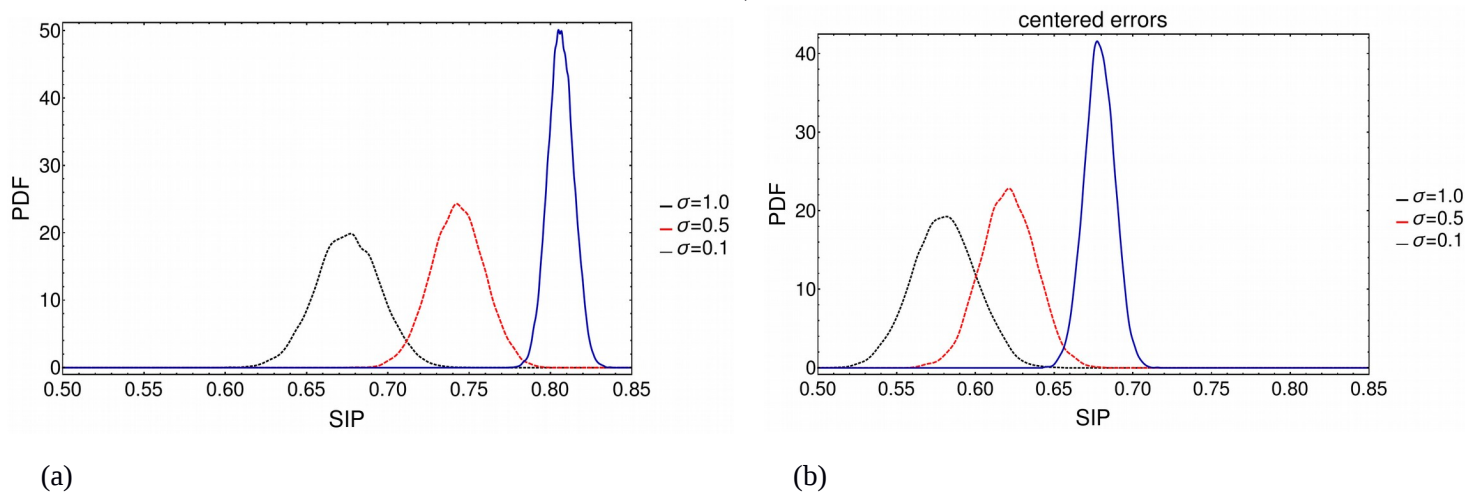

(a)                                                                                     (b)

*Figure 5. Distribution of the systematic improvement probability (SIP) for different uncertainties of the reference data (full curve: σ=0.1, dashed curve: σ=0.5, dotted curve, σ=1.0 eV): (a) errors; (b) centered errors.*

The reason for the displacement of the SIP distribution to lower values after increasing σ is due to the definition of SIP: we are looking at absolute errors. As σ in the reference data increases, the difference between LDA and HSE06 fades off. In fact, it can be shown that as long as $\sigma > e_{m,i}$, there is no distinction of the methods, the error staying determined by the uncertainty of the reference. The displacement would not have been present if we had considered signed errors.

## Conclusion

Benchmarks data for testing functionals are very useful. However, the usual measures provided, such that mean absolute errors, are not sufficient, and a more detailed study of the data sets is needed. In particular, the user should be able to input the expectations he has from the method, find out what the risks are, and decide whether he is willing to take them. A wider set of examples illustrating these points is presented in Ref.[7].

But also experimental data need to be provided with more care. In order to produce a reliable benchmark, reliable reference data are required, and for experimental data this means that error bars should be given, and estimates of errors from other sources (*e.g.*, coming from the models used in the treatment of the raw data).

## Acknowledgments

It is a pleasure to dedicate this paper to Yuri Grin, who always realized the importance of the interaction between theory and experiment.